# A Tabu Search based clustering algorithm and its parallel implementation on Spark

Yinhao Lu,   Buyang Cao[#]

School of Software Engineering

Tongji University, 4800 Cao'An Road, Shanghai, China 201804

Cesar Rego

School of Business Administration

University of Mississippi, University, MS 38677, USA.

Fred Glover

ECEE, College of Engineering & Applied Science

University of Colorado, Boulder, CO 80309, USA

***Abstract*** The well-known *K*-means clustering algorithm has been employed widely in different application domains ranging from data analytics to logistics applications. However, the *K*-means algorithm can be affected by factors such as the initial choice of centroids and can readily become trapped in a local optimum. In this paper, we propose an improved *K*-means clustering algorithm that is augmented by a Tabu Search strategy, and which is better adapted to meet the needs of big data applications. Our design focuses on enhancements to take advantage of parallel processing based on the Spark framework. Computational experiments demonstrate the superiority of our parallel Tabu Search based clustering algorithm over a widely used version of the *K*-means approach embodied in the parallel Spark MLlib system, comparing the algorithms in terms of scalability, accuracy, and effectiveness.

***Keywords***: Clustering,  *K*-means, Tabu Search, parallel computing, Spark, Big data

1. INTRODUCTION

The purpose of a clustering process is to group a set of (abstract or physical) objects into multiple classes, so that the objects in each class (cluster) are similar according to certain rules or criteria. A clustering algorithm in general seeks to build the clusters by the two interrelated criteria of selecting objects to lie in the same cluster that are as similar as possible while undertaking to assure that objects that lie in different clusters are as dissimilar as possible, where the definition of similarity can be problem dependent. Clustering problems can be found in applications ranging from data analytics to logistics applications as documented in the surveys of Jain et al. (1999), Berkhin (2002), Grabmeier and Rudolph (2002), Xu and Wunch (2005), and Jain (2010). The search for new clustering algorithms that best fit the different applications has proved fundamentally important to many recent advances in the domains of biology, genetics, medicine, business, engineering, and social science, among others.

# Corresponding author

Data-driven decision making as well as the burgeoning demand for data analytics has inspired increasing numbers of scholars as well practitioners to develop and apply clustering algorithms. Clustering is generally known as an unsupervised learning method (since no prior knowledge is provided that determines which objects should be grouped in a common cluster) and plays a crucial role in finding patterns and trends in the datasets. The role of clustering is highlighted in Grabmeier and Rudolph (2002), who propose a variety of criteria to evaluate the quality of clusters along with suggestions about how to select solution methodologies. From another perspective, Łuczak (2016) proposes a hierarchical clustering approach for classification problems involving time series datasets. Although Łuczak's method works well for classifying time series data, it cannot be applied directly to most other clustering problems without major modifications.

Data analytics based on clustering are especially pervasive in the public health arena. Glatman-Freedman et al. (2016) discuss the use of near real-time spatiotemporal cluster analysis to devise strategies for combatting some enteric bacteria diseases. With the help of this type of analysis, the source of a disease can be identified in a timely manner to enable appropriate measures to be taken before it becomes widespread. Clustering applications also abound in the field of logistics, where spatial and other relational restrictions often exist to limit the choice of objects that can lie in a common cluster. To solve the problems encountered in such applications, Cao and Glover (2010) present an algorithm based on Thiessen-polygons that proves highly effective in generating clusters that satisfy restrictions on balancing and connectivity. In this case, the entire service area for the logistics application is divided into several geographically connected and balanced subareas that account for geographic obstacles and constraints imposed by business logic, enabling a logistics service provider to offer better and more efficient services for customers.

Among various clustering approaches, the *K*-means clustering algorithm (MacQueen, 1967) is one of the most popular and widely applied. Modern data analytical solver packages, including open source packages such as *R* and *Spark MLlib*, include the *K*-means clustering algorithm among their offerings. Nevertheless, the *K*-means clustering algorithm exhibits some limitations that need to be addressed to solve clustering problems more effectively. Notably, the algorithm relies on an initial determination of a set of centroids in order to launch subsequent steps that assign objects to these centroids and thereby identify new centroids. The choice of these initial centroids has a major influence on the structure and quality of the final clusters produced, and hence an improper selection of these centroids can lead to unsatisfactory outcomes. Moreover, when starting from any given set of centroids, the *K*-means process for successively assigning objects and generating new centroids relies on a heuristic approximation of conditions necessary to achieve optimality, and the solution trajectory produced by this approach may not be particularly effective. Thus, researchers and practitioners have developed a variety of procedures in an attempt to overcome these limitations.

Xiong et al. (2011) utilize the exiting max-mix distance algorithm as the foundation for improving the *K*-means approach, to overcome the great fluctuations in the final outcomes produced by generating initial centroids randomly. This approach accelerates the convergence of the *K*-means algorithm but does not perform well in terms of *accuracy*. Lin et al. (2012) attempt to improve the *K*-means algorithm by integrating it with an approach based on particle swarm optimization and multiclass merging. Their experimental tests show that their approach yields better overall outcomes than the original *K*-means algorithm. In general, the solutions created by the *K*-means algorithm are influenced greatly by the initial solution settings. Poorly selected initial solutions lead to undesirable final solutions. To overcome this problem, Celebi et al. (2013) and Wang and Bai (2016) propose a variety of methods to generate better initial solutions rather than relying on randomly picked centroids. The goal of generating better initial solutions is also one of the motivations for our paper.

Furthermore, like other heuristics, the *K*-means algorithm suffers from its susceptibility to become trapped in local optima. To address this issue, for a data analytical project Cao et al. (2015) propose a Tabu Search algorithm that aims to produce tighter and more cohesive clusters based on incorporating criteria for these characteristics in the objective function. The algorithm succeeds both in obtaining better global solutions and in producing clusters that are more cohesive, but the computation time is greater than required by a well-implemented *K*-means method. To address the performance of *K*-means in solving large-scale problems, Bhimani et al. (2015) propose three parallel computational models including shared memory, distributed memory, and GPU-based heterogeneous computing. Computational experiments on image datasets showed these parallel versions of *K*-means to be 30 times faster on average than the sequential counterpart. To overcome the well-known sensitivity of *K*-means to the initial solution, a set of candidate solutions is first generated in parallel and the best is selected to initiate the algorithm. This approach proved essential to maximize algorithm speedup. Xu and Cao (2015) present a method for parallelizing a Tabu Search clustering algorithm utilizing a subspace partitioning principle. The computational results are appealing in terms of both solution quality and computing performance, but a gap remains in achieving outcomes that are ideal. The implementation utilizes a multi-core platform to run a multi-thread version of the Tabu Search clustering processes while employing a subspace partitioning principle to carry out the data transactions. However, the underlying algorithm structure is not compatible with a big data computational framework such as Spark.

With the existence of big data computing infrastructures including cloud computing, we are now increasingly able to analyze and process large volumes of data efficiently. However, a variety of classic optimization problems that require many iterations to solve have yet to benefit from these latest technologies. Recourse is sometimes made to heuristics that yield local optima of questionable quality. More ambitious metaheuristic based approaches yield better solutions but in some cases face a challenge to produce such solutions within a reasonable span of computation time. As more businesses move their IT services to centralized environments such as cloud platforms, it is necessary to develop and implement optimization algorithms more efficiently to accommodate the ever-increasing scale of practical problems. Consequently, we are motivated to explore the possibilities of utilizing big data computing infrastructures like Spark to solve large-scale optimization problems.

As one of a series of research projects, in this paper we address two major issues encountered in solving large-scale clustering problems, namely, the potentially poor quality of local optima obtained by simple clustering algorithms such as *K*-means, and the generally poor computational times produced by more complicated clustering algorithms. We propose a Tabu Search strategy to tackle the local optimality problem in conjunction with a Spark platform parallel processing implementation that makes it possible to handle large-scale problems more efficiently. The main contributions of the paper are:
- Design and implement the parallel mechanism for the algorithm to operate within a big data computing infrastructure.
- Develop a strategy for generating initial clustering solutions to provide stable and robust outcomes.
- Design the solution neighborhood structure and an associated candidate list selection strategy so that the solution procedure will be capable of effectively exploiting the MapReduce operations.
- Establish the merit and feasibility of applying metaheuristics such as Tabu Search within the Spark environment, thereby encouraging other researchers to explore the use of metaheuristics in big data environments to solve large-scale optimization problems.

The paper is organized as follows: Section 2 describes the clustering model, the Tabu Search based clustering algorithm, and its parallel implementation on Spark platform. Computational results are presented in Section 3. Finally, section 4 concludes the paper with our findings and future research directions.

2. MODEL AND ALGORITHMS

*2.1 The model*

In the following we represent similarity by a distance measure, and seek a collection of clusters that minimizes intra-cluster distance and maximizes inter-cluster distance. We call the objects to be clustered as data points and refer to the set of objects as a dataset. Consider a dataset of $N_p$ objects. Each data point in the dataset has $k$ attributes, i.e., it is $k$-dimensional. A data point $x_t$ will be represented by a vector $x_t = (x_{t1}, x_{t2}, ..., x_{tk})$. The underlying dataset then can be represented by

$$X = \{x_t : t = 1, ..., N_p\} \tag{2.1}$$

where $N_p$ identifies the total number of data points. Let $N_s$ be the total number of clusters to be built and denote each cluster by $C_i$, then the resultant cluster set will be:

$$C = \{C_i : i = 1, ..., N_s\} \tag{2.2}$$

The goal of our clustering problem is to group data points into a pre-defined number of clusters by the criteria previously discussed so that data points lying within the same cluster are as close (similar) as possible points lying in different clusters should be as far apart (dissimilar) as possible.

We employ the notation $Score(x_i, x_j)$ from the Tabu Search clustering paper of Cao et al. (2015) to represent the similarity of a pair of data points $(x_i, x_j)$. More precisely, $Score(x_i, x_j)$ describes the degree of correlation between two data points, so that smaller values indicate a greater desirability for assigning the points to the same cluster and larger values indicate a greater desirability for assigning them to different clusters. Here we suppose that Euclidean distance is used to describe the similarity of two data points in a $k$-dimensional space though other distance measures (including those that do not satisfy the definition of a norm) can also be used. Hence, for our present purposes we define

$$Score(x_i, x_j) = \sqrt{\sum_{p=1}^{k}(x_{ip} - x_{jp})^2} \tag{2.3}$$

Our algorithm makes use of distances between data points in several different ways depending on the objective, though in general we will want to identify a data point $x_i$ that maximizes or minimizes the sum of the distances from the data point to another data point or to a set of data points. In some instances, the objective may be to determine whether the sum of distances from a data point to a set of points fall below a specified threshold. The general function to perform those computations and identify the desired data point may be written as follows:

$$V(i) = \sum Score(x_i, x_j), \quad x_i \in L, x_j \in U, x_i \neq x_j \tag{2.4}$$

A main use for this function is to compute the total distance from all data points in a cluster to its centroid, which is also a data point of the cluster. Let $x_i$ denote the centroid of cluster $C_i$, the total distance (from all data points $x_j$ to the centroid $x_i$) of cluster $C_i$ is obtained by setting $L = \{x_i\}$ and $U = C_i \setminus \{x_i\}$.

Under the foregoing settings, the objective function of our clustering problem that we seek to minimize is then denoted by the following formula:

$$objVal = \sum_{i=1}^{N_s} V(i) \qquad (2.5)$$

*2.2 Tabu Search algorithm design*

To accommodate the parallel implementation of our algorithm, we solve clustering problems by applying the so-called *centroid-driven* approach. Unlike the regular *K*-means algorithm where centroids (except for the initial centroids) are recalculated after data points are shuffled, we find better centroids for all clusters and then assign data points to the proper clusters to optimize the value of $objVal$ defined in (2.5). As long as cluster centroids are defined, we can apply the same logic used in the *K*-means algorithm to identify the clusters associated with these centroids simply by assigning each point to the centroid closest to it. As we show in detail in the next section, this strategy is easily parallelized.

The key strategies of the Tabu Search component of our clustering algorithm may be described as follows. Tabu Search (TS) (Glover 1989, 1990) is a metaheuristic algorithm designed to guide subordinate heuristic search processes to escape the trap of local optimality. TS is distinguished from other metaheuristics by its focus on using adaptive memory and special strategies for exploiting this memory. Memory is often divided into short-term and long-term memory, and the Tabu Search strategies for taking advantage of this memory are often classified under the headings of intensification and diversification.[1] A common form of TS short-term memory is a recency-based memory that operates to temporarily prevent recently executed moves from being reversed for a duration (number of iterations) known as the tabu tenure. Other types of short-term and long-term memory make use of frequency-based memory as described in Glover and Laguna (1997). In our current implementation we make use of a simple version of TS that uses recency-based memory alone.

---

[1] The intensification/diversification terminology introduced in Tabu Search has subsequently been adopted by many other metaheuristics.

Intensification strategies in TS are designed to focus the search more strongly in regions identified by past search history and by current evaluations as likely to harbor good solutions, while diversification strategies focus the search more strongly on visiting regions that have not been examined before. These two strategies are interdependent, and the best forms of each result by including reference to the goals of the other.

We employ the Spark platform in this setting to take advantage of the fact that Spark has become a standard platform for processing and analyzing large datasets. The parallel version of our algorithm is chiefly based on the fact that the objects to be clustered do not impact each other during the step in which they are reallocated to new centroids to create new clusters.

Following standard terminology, the transition from a current solution to a new one is called a *move*. Utilizing the centroid-driven idea, the type of move we exploit by parallelization consists of (a) selecting a data point as the new centroid of a cluster, and (b) reassigning data points to their closest new centroids to create corresponding new clusters. However, we modify (b) by taking account of the objective function value $objVal$ defined in (2.5) as a basis for generating improved clusters.

**Neighborhood determination**: a neighborhood is the solution subspace of the current solution which defines the available moves for generating a new solution at the next iteration. Neighborhood design and construction is highly important for an efficient Tabu Search algorithm. An inappropriate neighborhood may miss the opportunity to explore more promising solution spaces or may result in spending too much time examining unnecessary or unpromising spaces. We employ a neighborhood definition that results by creating a sphere that places a centroid at its center. For a given cluster $C_i$ with centroid $x_i$, we define the neighborhood $N(i)$ of this cluster as follows:

$$N(i) = \{x_j : |x_j - x_i| \leq R_i, x_i, x_j \in C_i, x_i \neq x_j\} \qquad (2.6)$$

Here $R_i$ is the radius of the neighborhood sphere, and $|x_j - x_i|$ is defined by (2.3). As mentioned earlier, we employ the convention whereby a centroid itself is treated as a data point. Any data point that lies in cluster $C_i$ and satisfies the conditions indicated in (2.6) becomes an element of the neighborhood $N(i)$, and hence is a candidate to become a new centroid at the next iteration. Thus, using (2.6), we first construct a neighborhood sphere from the centroid of a current cluster, and then select an element in this sphere to be the centroid of a new cluster. Then, a neighboring solution is obtained by reassigning all remaining elements (data points) to their closest centroid. Once this process is completed for all data points other than the centroids in current clusters, the highest evaluation neighboring solution is chosen to define the new clusters for the next round. In other words, a *move* is performed by letting the best solution in the current neighborhood become the new working solution. This process defines one iteration of our local search procedure on which Tabu Search is superimposed to provide appropriate guidance, as explained later.

The sphere radius value plays an important role in our algorithm search strategy and is critical to performance. The following procedure determines the radius value $R$. Relative to a given cluster $C_i$ with centroid at $x_i$ and the data points, $R_i$ is defined as follows:

$$R_i = \sum_{i=1}^{|C_i|} Score(x_j, x_i)/|C_i|, x_i, x_j \in C_i, x_i \neq x_j \qquad (2.7)$$

Clearly $R_i$ varies as a function of the individual cluster $C_i$ it is associated with, and in this sense changes dynamically for each cluster as the cluster's composition changes. The overall neighborhood of the Tabu Search algorithm is then:

$$NB = \bigcup_{i=1}^{N_s} N(i) \qquad (2.8)$$

**Tabu list**: a tabu list holds the information for those solutions that cannot be revisited during the next *t* iterations. The solutions referenced by the tabu list are called *tabu*, and *t* designates the *tabu tenure*. The fact that in our algorithm clusters are formed by assigning each data point to its closest centroid allows for a move to be completely defined by two data points in a common cluster that swap their labels (from centroid to non-centroid and vice versa). More formally, let $O = \{o_1, \dots, o_{N_s}\}$ be the set of current centroids in a solution and $\bar{O} = X \setminus O$ be its complement relative to data set $X$. Let $x_j$ and $x_k$ be two data points in the current cluster $C_i$ where $x_j \in O$ and $x_k \in \bar{O}$. A move is then defined by setting $o_i = x_k$, which automatically kicks $x_j$ out of set $O$; hence, after swapping labels we have $x_k \in O$ and $x_j \in \bar{O}$. The corresponding neighboring solution is then obtained by assigning each data point in $\bar{O}$ to its closest data point in $O$. For the purpose of short term memory guidance, it is sufficient to impose a restriction that prevents $x_j$ from moving back into set $O$ during an appropriate number of iterations (from the current one) which we denote by $t$. The tabu list is implemented as a linear array $TL(j)$, $j = 1, \dots, N_p$, with each component being the iteration number at which the tabu restriction on $x_j$ is relaxed. Whenever an element $x_j$ leaves the centroid set $O$, its tabu status is set to $I + t$, where $I$ denotes the current iteration. To provide greater flexibility and an opportunity to find better solutions, the tabu status of a solution can be overridden (or lifted) if a so-called *aspiration criterion* is met. In our application, the aspiration criterion is the most basic one, which is satisfied if the solution produced by the move is better than the best solution found so far. In the section for computational experiments we provide more details for how to set the tabu tenure for the clustering problems to be solved.

**Candidate list**: a candidate list $CL$ is a subset of the neighborhood $NB$ which is generated to reduce the computational effort of examining the complete neighborhood, using a design that focuses on moves that are anticipated to be the more promising ones for uncovering improved solutions. We use the following process to pick data points to become members of the candidate list:

___

//**Candidate List Creation**
Procedure CeateCandidateList (
    $C_i$ // the $i^{th}$ cluster
    $N(i)$ // the component neighborhood for cluster $C_i$
    $R_i$ // the radius defined by (2.7) for cluster $C_i$
    $V(k)$ // sum of distances (scores) for data point $x_k$ determined by (2.4)
    $V'(k)$ // the value of $V(k)$ obtained at the most recent local optimum
    $n(i)$ // $< |N(i)|$, the number of elements (data points) to be included in the candidate list $CL(i)$ for cluster $C_i$
    $\varepsilon$ // the number of data points whose current $V(k)$ values are less than $V'(k)$, which could be $0$
    $\delta$ // an integer drawn randomly from the interval $[0, |N(i)| - \varepsilon]$
)
//creating component neighborhood
1    For all clusters $C_1, C_2, \ldots, C_s$
        Create $N(i)$ using radius $R_i$
//computing total distance (sum of scores)
2    For all clusters $C_1, C_2, \ldots, C_s$
3        $i =$ the current cluster index
4        $x_i =$ the centroid of cluster $C_i$
5        For each point $x_k \in N(i)$ other than $x_i$
6           $L = \{x_i\}$, $U = N(i) \backslash \{x_i, x_k\}$
7           Computing $V(k)$ using (2.4)
//creating candidate list
8    Sort the data points $x_k \in N(i)$ $(i = 1, \ldots, N_s)$ in ascending order of their $V(k)$ values
9    For all clusters $C_1, C_2, \ldots, C_s$
10      $n(i) = \varepsilon + \lambda$
11      $CL(i) =$ first $n(i)$ data points in the sorted $N(i)$ list

12    $CL = \cup_1^{N_s} CL(i)$

13    return $CL(i), i = 1, \ldots, N_s$
___

**Intensification and diversification strategies**: As previously noted, the intensification strategy in Tabu Search guides the search to explore more attractive regions of the solution space while the diversification strategy encourages the search to explore rarely examined regions. In our implementation we focus chiefly on speed of execution and therefore use extremely simple types of intensification and diversification strategies.

Our intensification strategy consists precisely of restricting attention to members of the candidate list to pick new centroids for clusters, given the design of $CL$ which focuses on the highest evaluation moves. Correspondingly, the diversification strategy merely consists of using the entire neighborhood $NB$ as the candidate list without restriction to the higher evaluation moves. Since the candidate list for the diversification strategy is obviously larger than the one for the intensification strategy, we thereby gain the possibility of exploring regions that are rarely examined by the intensification strategy. Evidently, more sophisticated intensification and diversification strategies are possible, but we find that these simple approaches perform in a satisfactory manner to support our parallel implementation.

**Stopping criteria**: our clustering algorithm terminates if one of the following two conditions is met, as detailed in Section 2.4:
- A predefined maximum number of iterations is reached,
- An improvement is not found in two consecutive calls of the Tabu Search procedure (one for the intensification and the other for diversification).

*2.3 Construction of initial centroids*

Our method of constructing an initial solution is based on a maximum distance idea expressed in terms of a sum of distances from existing centroids. Unlike *K*-means, the centroids in our algorithm are real data points. We assume the number of clusters, $N_s$, is greater than 1. Initial solutions are created as follows:

---

//**Initial Solution Creation**
Procedure CreateInitialSolution (
    $C_i$ // the $i^{th}$ cluster
    $N_s$ // the number of clusters to be created
    $V(k)$ // sum of distances (scores) for data point $x_k$ determined by (2.4)
    $L$ // the set of centroids of all clusters
)
//selecting first centroid
1    Randomly select a data point $x_j$ as the first centroid
2    $L = \{x_j\}$
3    $U = X \setminus \{x_j\}$.
//creating the centroids for the rest clusters
4    Repeat the following steps until $N_s$ centroids are created

```
5       maxV = -1
6       nextCentroid = null
//selecting the next centroid
7       For all data point $x_k \in U$
8           $V(k)$ = sum of distances (scores) to all exiting centroids in $L$ upon (2.4)
9           If $V(k) > maxV$
10              $maxV = V(k)$
11              $nextCentroid = x_k$
12      $L = L \cup \{nextCentroid\}$, $U = U \setminus \{nextCentroid\}$
// creating initial clusters
13   Assigning $x_k \in U$ to its closest centroid $x_i \in L$ to expand cluster $C_i$
14   return $C_i, i = 1, \dots, N_s$.
```
___

Again, we have focused on simplicity rather than sophistication. A variety of more advanced procedures are given in Glover (2016) that replace the maximum distance measure with generalizations of a MaxMin distance measure utilizing iterative refinement and adaptive thresholds.

Figure 2.1 depicts the process of creating the initial centroids for three clusters. Point *X* which is the first centroid is selected randomly. Then we select the point farthest from *X*, identifying point *Y* as the second centroid. The third centroid is chosen to be the point possessing the largest sum of distances to the previously selected centroids, hence yielding the point *Z*. The remaining data points are then assigned to their closest centroids to form the initial clusters. We see that unlike the traditional *K*-means algorithm in which the centroid of a cluster is the center of gravity, our initialization algorithm selects centroids of clusters which are data points. This likewise constitutes an instance of a more sophisticated type of strategy in the pseudo-centroid clustering approach of Glover (2016).

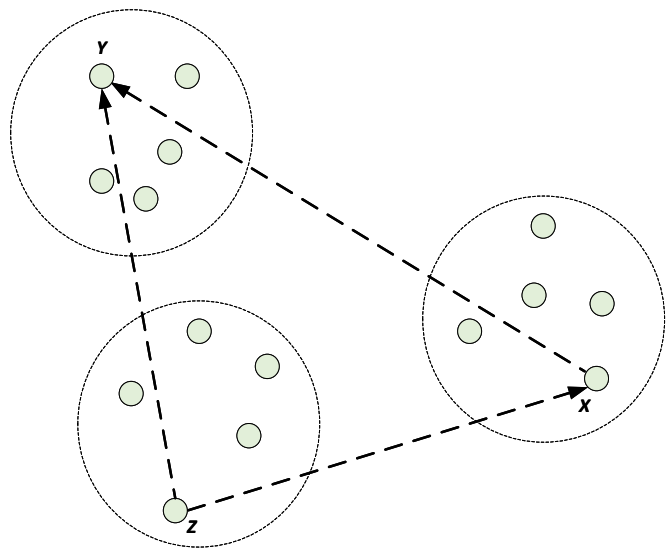

**Figure 2.1 Construction of initial centroids**

*2.4 Tabu search based clustering algorithm*

We assemble the components described above to produce our complete algorithm as follows:
______________________________________________________________________________

**//Tabu Search based clustering algorithm**
Procedure TS_Clustering (
    $S$ // the solution at the current iteration
    $S'$ // the locally best solution at the current iteration
    $S^*$ // the best solution found so far
    $Z(.)$ // the objective function value (2.5) for a solution "."
    $NI$ // the number of consecutive iterations without improvement of $S^*$
    $MaxNI$ // a predefined maximum limit on $NI$
    $N_s$ // the number of clusters to be created
    $X$ // the data (points) set
)
//constructing initial solution
1    Calling *CreateInitialSolution*() and getting initial solution $S = \{C_i : i = 1, \dots, N_s\}$
2    $NI = 0$
3    Setting corresponding values $Z(.)$
// Tabu search procedure
4    For all clusters $C_1, C_2, \dots, C_s$
5        $CL(i) = CreateCandidateList(i)$ // creating candidate list upon cluster index $i$
6    For all candidate lists $CL(1), CL(2), \dots, CL(N_s)$
7        For each data point $x_k \in CL(i)$
8            Setting $x_k$ to be the new centroid
9            $S =$ assigning non-centroid data points to theirs closest centroids
10          If ($S$ meets tabu condition and $Z(S) < Z(S')$) or $Z(S) < Z(S^*)$
11            $S' = S$
12            $Z(S') = Z(S)$
13    If $Z(S') < Z(S^*)$
14       $S^* = S'$
15       $NI = 0$
16    Else
17       $NI = NI + 1$
18    If $NI < MAXNI$
19       $S = S'$
20       Goto 4
21    If both intensification and diversification executions of Steps 4–20 have failed to improve $S^*$
22       Terminating the algorithm
23    Else
24       Switching either to diversification or intensification strategy upon the current one
25       $NI = 0$
26       Goto 4
______________________________________________________________________________

*2.5 Parallelization*

By utilizing latest advances in computer hardware and computing frameworks, the parallelization of algorithms has been shown to be useful for solving large-scale data analytical and optimization problems related to clustering in Rego (2001), Xu and Cao (2015), Gopalani and Arora (2015), and Wang et. al. (2015). The motivation for parallelizing our algorithm derives from the fact that the data points are not affected by each other when they are reassigned to the closest centroids. The process of reallocating data points to generate new clusters can be parallelized by using the *mapPartition* transform in Spark. At the end of the *mapPartition* phase, the selection of new centroids for the resulting clusters can also be treated as independent due to our special neighborhood design. In this case, the evaluation for each cluster centroid can be implemented in parallel using the Spark *reduce* operation.

There are multiple Map-Reduce operations for the parallel version of our algorithm where the entire dataset will be divided into several blocks, and each Map or Reduce is responsible for the exploration of a data block. Let $N_{mr}$ be the number of Map-Reduce operations, each of which is associated with one computing unit in the parallel computing environment. When the whole dataset is split between these operations and processed simultaneously, the computation time is expected to be reduced significantly.

We parallelize two components of our algorithm. The first component is the process of reallocating/reassigning data points to the clusters based on assigning them to the closest centroids. Since the data points can be treated independently in this process, during the map phase each map assigns data points to their nearest centroids in parallel. The second component is the process of updating the centroid of each cluster. Since the neighborhood creation is based on the sphere of the current cluster, the selection of the best centroids for individual clusters does not create any interference between the clusters. This step can be parallelized as well.

The reduce process relies on the output of map: <centerId, pointList>. By referring to centerId we can obtain all associated data points forming the cluster, calculate the sphere radius of the current cluster by formula (2.7), and obtain the candidate list from the neighborhood sphere for either the intensification or diversification strategy. At each iteration, multiple map-reduce operations are run simultaneously on the splitting datasets and we choose the best solution by merging all solutions from the individual map-reduce processes for the current iteration.

*2.6  Parallel implementation on Spark*

Based on the discussion of the preceding section, the parallel version of our algorithm is quite like the non-parallel version with the addition of customization to accommodate the map-reduce operations. We will omit some details in the description of the parallel version that have been included in the non-parallel algorithm.

______________________________________________________________________

//**Core Parallelized Tabu Search based clustering algorithm** (showing master and slave control blocks and Spark methods)

Procedure Core_Parallel_TS_Clustering (

    $S$ // the solution found

)

//core parallel Tabu search

**Master:**

1    Calling *data.mapPartitions* (Spark method) to split the dataset into $N_{mr}$ parallel blocks

2    Each block gets a portion of the data set together with the centroids

**Slave (Parallel):**

3    Assigning each data point to its closest centroid to form partial clusters $PC_1, PC_2, \ldots, PC_s$

4    For all partial clusters $PC_1, PC_2, \ldots, PC_s$

5        Computing $PV(i)$ // the value defined by (2.4) for the partial cluster

**Master:**

6    $S = $ merging partial clusters from all slaves by calling

        *reduceByKey(getNewCenters(data))*

7    Return $S$

______________________________________________________________________

The overall algorithm can be written as follows:

______________________________________________________________________

// **Parallelized Tabu Search based clustering algorithm** (showing master and slave control blocks and Spark methods)

Procedure Parallel_TS_Clustering (

    $N_{mr}$ // the number of Map-Reduce operations

    $RDD$ // the Resilient Distributed Dataset holding data points to be clustered

    $S'$ // the locally best solution at the current iteration

    $S^*$ // the best solution found so far

    $Z(.)$ // the objective function value (2.5) for a solution "."

    $X$ // the data (points) set

    $Com\_Centroids$ // all feasible centroid combinations of candidates picked from $CL(i)$

    $Com\_Centroid(q)$ // one element in $Com\_Centroid$

    $Centroid\_Set$ // the set of centroids of all clusters

)

//constructing initial solution

**Master:**

1    Loading data points $X$ as a $RDD$

2    Performing steps 1– 6 of *CreateInitialSolution*() to generate centroids

//calling core algorithm

3    S = *Core_Parallel_TS_Clustering*()

4    Updating $S'$, $S^*$, and $Z(.)$ similar to steps 10 – 17 of *TS_Clustering*()

5   Perform the following steps until stopping criterion is met
6       Grouping $C_1, C_2, \ldots, C_s$ into $N_{mr}$ blocks
**Slave:**
        //creating the candidate list upon cluster index $i$ in parallel
7       $CL(i) = CreateCandidateList(i)$
**Master:**
8       *Com_Centroids* = the set of feasible centroid yielded upon $L(i)$, $i = 1, \ldots, N_s$.
9       For each *Com_Centroid(q)* in *Com_Centroids*
10          *Centroid_Set* = *Com_Centroid(q)*
11          $S = Core\_Parallel\_TS\_Clustering()$
12          Updating $S'$, $S^*$, and $Z(.)$ similar to steps 10 – 17 of *TS_Clustering()*
13  Return $S^*$

*2.7 Time Complexity Analysis*

In this analysis, we omit the time for computing distances between all data pairs since it is a constant. Nevertheless, in our computational experiments all distances are calculated on the fly. According to the discussion in Section 2.6, the running time of the algorithm depends on the following factors:
- the number of iterations $N_l$,
- the number of clusters $N_s$,
- the number of data points $N_p$, and
- the number of parallel data blocks $N_{mr}$.

Because our computational framework splits the dataset into $N_{mr}$ blocks, the quantity of the data in each block or computing unit is approximately $N_p/N_{mr}$. In each parallel computing unit, the task of the mapPartition function is to assign data points to the corresponding clusters. For each data point we need to sort the distances from each data point to all $N_s$ centroids in ascending order. The quick sorting complexity is $O(N_s log(N_s))$, and the time to pick the shortest distance after sorting is negligible. Therefore, the time complexity of this process is $O(N_s log(N_s) N_p/N_{mr})$.

The reduceByKey function first determines the radius of the neighborhood sphere for each cluster. For each cluster the corresponding computational time is proportional to the number of data points involved in each computing unit. Therefore, the time complexity of determining $R_i$ will be $O(N_p/N_{mr})$ and the time complexity for all clusters is $O(N_s N_p/N_{mr})$. To determine the candidates from a neighborhood we compute the cost $V(j)$ for all $x_j$ in the neighborhood and sort them in ascending order. Assuming the number of data points in the neighborhood is $\alpha$, the time complexity of these two operations is $O(\alpha(1 + log(\alpha)))$. The value of $\alpha$ is about an $N_p/2N_{mr}$ order of magnitude operation by the neighborhood definition, and the time complexity of determining candidates for all clusters is

$$O\left(\frac{N_s N_p}{2N_{mr}}\left(1 + log\left(\frac{N_p}{2N_{mr}}\right)\right)\right)$$

.

If the algorithm runs for $N_l$ iterations, the overall time complexity for the parallel version (implemented in the Spark framework) is therefore:

$$O\left(\frac{N_s N_p}{N_{mr}}\left(N_l(1 + log(N_s)) + \frac{1}{2}\left(1 + log\left(\frac{N_p}{2N_{mr}}\right)\right)\right)\right)$$

The time complexity of the traditional *K*-means algorithm is $O(kN_l N_s N_p)$ if the data points are $k$ dimensional. The time complexity of our algorithm is similar to that of the *K*-means algorithm with extra time spent on creating neighborhoods and candidate lists.

3. EXPERIMENTS

We have implemented the algorithm presented in section 2 on the Spark platform using a computational environment consisting of four machines, which are divided into one master node and three slave nodes. The hardware configuration of each machine is the same: CPU Intel core2 2.2GHZ, RAM 2GB, Hard Disk 500GB, Ethernet 100M/s. The overall architecture is depicted as follows:

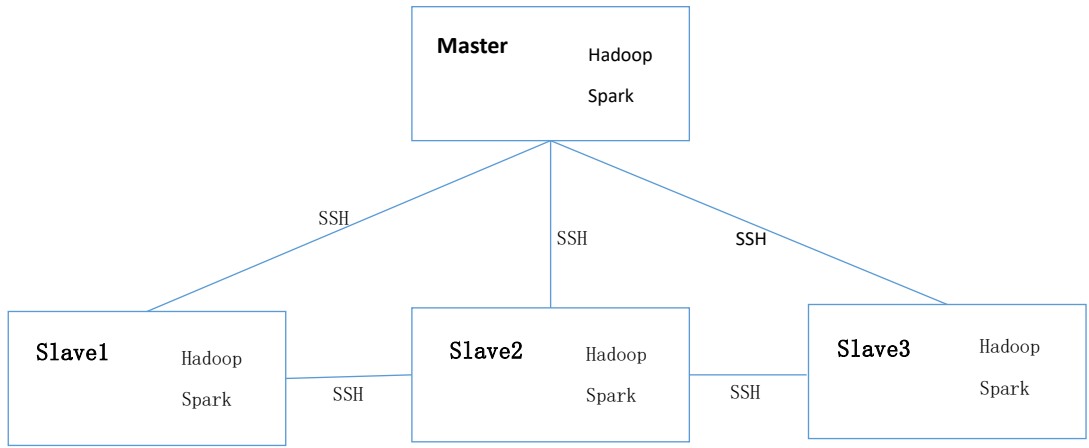

**Figure 3.1 Overall architecture**

The datasets (Iris, Wine, Yeast, and Seeds) for the computational experiments are downloaded from the UCI open dataset depository (https://archive.ics.uci.edu/ml/datasets). The data types contained in the datasets are different, so that we may validate the applicability and performance of our algorithm. Furthermore, all data points contained in the datasets have been labeled so that the correctness of the clustering outcomes can be verified relatively easily.

*3.1 The impacts of the algorithm parameters*

As previously mentioned, the tabu tenure has a significant impact on the quality of the results. We have used fixed tabu tenures instead of dynamic tenures to keep our algorithm simple, though dynamic tenures may generally be expected to produce better outcomes. Solution quality can also of course be influenced by the number of iterations permitted for carrying out the search process. To find reasonable algorithm parameters, we conducted preliminary computational experiments using different tabu tenures and numbers of iterations. The experiments provide some insights into selecting proper parameters for obtaining more satisfactory solutions. Let $N_{correct}$ denote the number of data points assigned to correct clusters and let $N_{total}$ denote the total number of data points. Then the quality of a solution is determined by the accuracy $P$ defined as follows:

$$P = \frac{N_{correct}}{N_{total}} \quad (3.1)$$

The original datasets contain labels for each data point identifying the cluster to which it should be assigned as a basis for identifying $N_{correct}$. The entries in the "Accuracy" column of all following tables is computed based upon this formula. The computational times listed in all tables include times not only for the algorithm and distance calculations but also the ones spent particularly in a distributed system such as task scheduling, job tracking and monitoring, data shuffling between different nodes, etc.

The neighborhood $N(i)$ described above plays an important role in obtaining satisfactory clustering results. To investigate the impacts of various neighborhood sizes in more detail, we tested the following three settings for the neighborhood size:

- **Small**: half of that defined in (2.7), i.e., $R_i/2$ for individual cluster $C_i$.
- **Standard**: as defined by (2.7), i.e., $R_i$ for all clusters $C_i$.
- **Large**: the largest distance among those from the current centroid to all data points. In this case, the neighborhood contains all data points in the current cluster. Note, however, that $CL$ in general is different from the neighborhood.

The best tradeoff between neighborhood size and number of iterations is key to maximum algorithm performance. On one hand, for constant neighborhood size more iterations will necessarily increase the computational time while less iterations may compromise solution quality. On the other hand, a similar result is expected when varying the neighborhood for constant number of iterations. The sensitivity of the algorithm to the tabu tenure should also be investigated under the various neighborhood sizes and numbers of iterations.

To gauge the impact of the algorithm parameters on performance, we conduct a set of experiments with different parameter settings, running the algorithm ten times for each instance and averaging the results. The outcomes are listed in Table 3.1.

**Table 3.1 The impacts of algorithm parameters**

| Dataset | Number of iterations | Tabu tenure | Radius size | Comp. Time (sec.) | Accuracy |
|---------|----------------------|-------------|-------------|-------------------|----------|
| Iris | 500 | 5 | small | 2.7134 | 0.6122 |
| Iris | 500 | 5 | standard | 3.3451 | 0.6928 |
| Iris | 500 | 10 | standard | 3.6190 | 0.6533 |
| Iris | 500 | 5 | large | 4.2296 | 0.7012 |
| Iris | 1000 | 5 | small | 4.3796 | 0.6384 |
| Iris | 1000 | 6 | small | 4.5120 | 0.6264 |
| Iris | 1000 | 8 | small | 4.5329 | 0.6211 |
| Iris | 1000 | 3 | small | 4.2198 | 0.6311 |
| Iris | 1000 | 2 | small | 4.1146 | 0.5821 |
| Iris | 1000 | 5 | standard | 8.0121 | 0.9234 |
| Iris | 1000 | 6 | standard | 8.0411 | 0.9125 |
| Iris | 1000 | 7 | standard | 8.0982 | 0.9052 |
| Iris | 1000 | 10 | standard | 8.1157 | 0.9021 |
| Iris | 1000 | 2 | standard | 7.7782 | 0.9192 |
| Iris | 1000 | 3 | standard | 7.8934 | 0.9203 |
| Iris | 1100 | 5 | standard | 8.5382 | 0.9234 |
| Iris | 1500 | 5 | standard | 11.2580 | 0.9234 |
| Iris | 1000 | 5 | large | 13.568 | 0.9255 |
| Iris | 1000 | 6 | large | 13.671 | 0.9203 |
| Iris | 1000 | 10 | large | 14.107 | 0.9128 |
| Iris | 1000 | 20 | large | 15.217 | 0.8972 |
| Wine | 500 | 5 | standard | 5.7653 | 0.5879 |
| Wine | 500 | 6 | standard | 5.7721 | 0.5823 |
| Wine | 500 | 10 | standard | 5.8283 | 0.5608 |
| Wine | 500 | 4 | standard | 5.7610 | 0.5822 |
| Wine | 1000 | 5 | standard | 11.8379 | 0.7239 |
| Wine | 1000 | 6 | standard | 12.2445 | 0.7157 |
| Wine | 1000 | 10 | standard | 12.4589 | 0.7012 |
| Wine | 1000 | 4 | standard | 12.2234 | 0.7211 |
| Wine | 1000 | 2 | standard | 12.1289 | 0.7067 |
| Wine | 1000 | 5 | large | 21.250 | 0.7311 |
| Wine | 1000 | 7 | large | 22.4329 | 0.7288 |
| Wine | 1000 | 2 | large | 20.1982 | 0.7215 |
| Wine | 1100 | 5 | standard | 14.1053 | 0.7239 |
| Wine | 1500 | 5 | standard | 16.9110 | 0.7239 |

The table shows that tabu tenures between 3 and 6 yield satisfactory solutions while tabu tenures outside of this range cause the solution quality to deteriorate. As may generally be expected, when the tabu tenure is set too small, cycling can occur and the process becomes trapped in a local optimum. Inversely, when the tabu tenure is too large, the neighborhood of admissible moves can become too restricted and prevent the method from discovering some of the higher quality solutions.

For a given tabu tenure and number of iterations, the small neighborhood sizes have a significant negative impact on the final results while standard and large neighborhood sizes produce almost the same outcomes. When we use a small neighborhood size, the candidate lists have fewer choices for their elements and eventually miss some opportunities to yield better solutions. On the other hand, the larger neighborhood sizes that involve consideration of a greater number of alternatives for building clusters will obviously need more time to evaluate (see the above time complexity and computational time analyses). In terms of computational time, on average the large neighborhood size needs more computational time than the standard neighborhood one does while the qualities of both settings are almost the same.

Table 3.1 also shows that running the algorithm for more than 1000 iterations does not yield significant benefit. On the other hand, running for less than 1000 iterations usually leads to poor solutions.

Based on our experiments, we set the tabu tenure to 5, the neighborhood size to be the standard, and the number of iterations to be 1000 respectively for all subsequent computational testing.

*3.2 Comparisons in Accuracy and Stability*

To compare the results obtained by our algorithm and the K-means algorithm of Spark MLlib we used the parameter settings indicated in section 3.1 by selecting the tabu tenure and the maximum number of iterations to be 5 and 1000 respectively for all four datasets. Since our algorithm chooses the first centroid randomly and the Spark MLlib $K$-means algorithm picks all initial centroids randomly, we run both algorithms multiple times to evaluate their overall accuracies and stabilities. The following four tables show the means and standard deviations of the solution accuracies.

Table 3.2 Accuracy of both algorithms on the Iris dataset

| The N-th experiment | Comp. time of K-means (s) | Accuracy of K-means | Comp. time of our algorithm(s) | Accuracy of our algorithm |
|---|---|---|---|---|
| 1 | 5.2341 | 0.8633 | 8.0901 | 0.9267 |
| 2 | 5.8921 | 0.8456 | 8.538 | 0.9267 |
| 3 | 6.3327 | 0.8698 | 7.8762 | 0.9223 |
| 4 | 6.5255 | 0.8321 | 8.1237 | 0.9116 |
| 5 | 7.1782 | 0.9139 | 8.3621 | 0.9267 |
| 6 | 5.9120 | 0.8140 | 8.0328 | 0.9233 |
| 7 | 6.2569 | 0.8569 | 8.2712 | 0.9187 |
| Average value | 6.1902 | 0.8565 | 8.1849 | 0.9223 |
| Standard deviation | 0.5598 | 0.031723988 | 0.2052 | 0.00557567 |

Table 3.3 Accuracy of both algorithms on the Wine dataset

| The N-th experiment | Comp. time of K-means(s) | Accuracy of K-means | Comp. time of our algorithm(s) | Accuracy of our algorithm |
|---|---|---|---|---|
| 1 | 8.1626 | 0.6836 | 12.014 | 0.7246 |
| 2 | 9.5498 | 0.7145 | 11.781 | 0.7233 |
| 3 | 8.6745 | 0.5166 | 11.216 | 0.7246 |
| 4 | 7.3589 | 0.6731 | 12.197 | 0.7242 |
| 5 | 5.7622 | 0.6389 | 11.009 | 0.7137 |
| 6 | 7.3561 | 0.6977 | 10.253 | 0.7244 |
| 7 | 5.8458 | 0.7012 | 11.833 | 0.7181 |
| Average value | 7.53 | 0.6608 | 11.472 | 0.7218 |
| Standard deviation | 1.2989 | 0.068095325 | 0.6337 | 0.004276625 |

Table 3.4 Accuracy of both algorithms on the Yeast dataset

| The N-th experiment | Comp. time of K-means(s) | Accuracy of K-means | Comp. time of our algorithm(s) | Accuracy of our algorithm |
|---|---|---|---|---|
| 1 | 9.1216 | 0.5233 | 11.115 | 0.6096 |
| 2 | 9.3110 | 0.5387 | 11.435 | 0.6177 |
| 3 | 7.1103 | 0.5899 | 10.329 | 0.6211 |
| 4 | 9.2352 | 0.6211 | 12.188 | 0.6185 |
| 5 | 8.4460 | 0.4910 | 11.649 | 0.6201 |
| 6 | 9.8320 | 0.5529 | 11.587 | 0.6195 |
| 7 | 8.9561 | 0.4967 | 12.172 | 0.6237 |
| Average value | 8.8589 | 0.5448 | 11.4964 | 0.6186 |
| Standard deviation | 0.8106 | 0.047659941 | 0.5950 | 0.004418522 |

**Table 3.5 Accuracy of both algorithms on the Seeds dataset**

| The N-th experiment | Comp. time of K-means(s) | Accuracy of K-means | Comp. time of our algorithm(S) | Accuracy of our algorithm |
|---|---|---|---|---|
| 1 | 5.8223 | 0.8826 | 7.3171 | 0.9467 |
| 2 | 6.4570 | 0.9105 | 8.2981 | 0.9345 |
| 3 | 4.8901 | 0.8944 | 6.3312 | 0.9488 |
| 4 | 5.1018 | 0.8367 | 7.2891 | 0.9488 |
| 5 | 3.8790 | 0.9533 | 7.3242 | 0.9391 |
| 6 | 5.2740 | 0.9269 | 8.2912 | 0.9431 |
| 7 | 5.1451 | 0.9102 | 6.3341 | 0.9502 |
| Average value | 5.2256 | 0.9021 | 7.7321 | 0.9445 |
| Standard deviation | 0.7389 | 0.03671818 | 0.7416 | 0.00585117 |

Our algorithm unsurprisingly takes longer computational time than the K-means algorithm embedded in Spark MLib because of the more sophisticated procedures introduced in our method. According to the last two rows of Tables 3.2 to 3.5, we can conclude that our algorithm provides a significant improvement over the K-means algorithm in terms of accuracy and robustness. We conjecture two reasons to account for this outcome. First is that the creation of initial solutions based on maximum distances can generate relatively robust initial solutions which facilitate the search process. Second is that the Tabu Search mechanism embedded in our algorithm can overcome local optimality more effectively and provide a better exploration of the solution space.

For the sake of clarity, we display the means and standard deviations of the accuracies listed in the preceding four Tables in Figures 3.2 and 3.3 respectively.

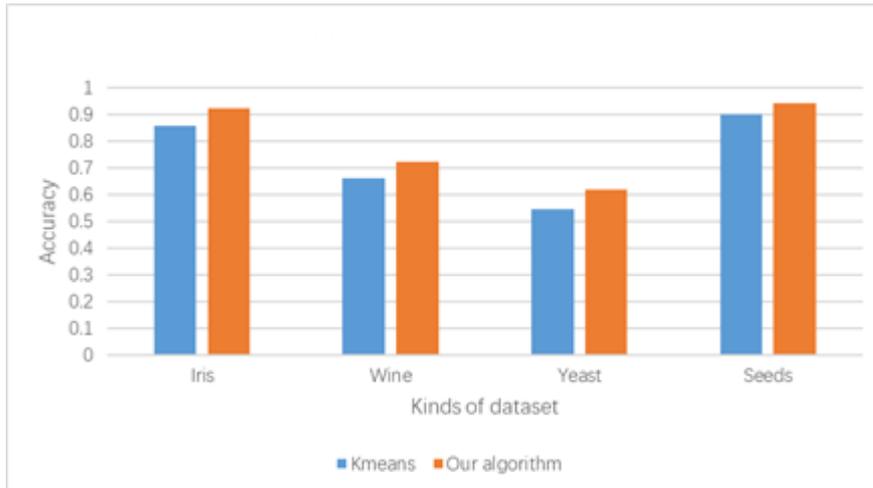

**Figure 3.2 Average accuracy of two algorithms on four datasets**

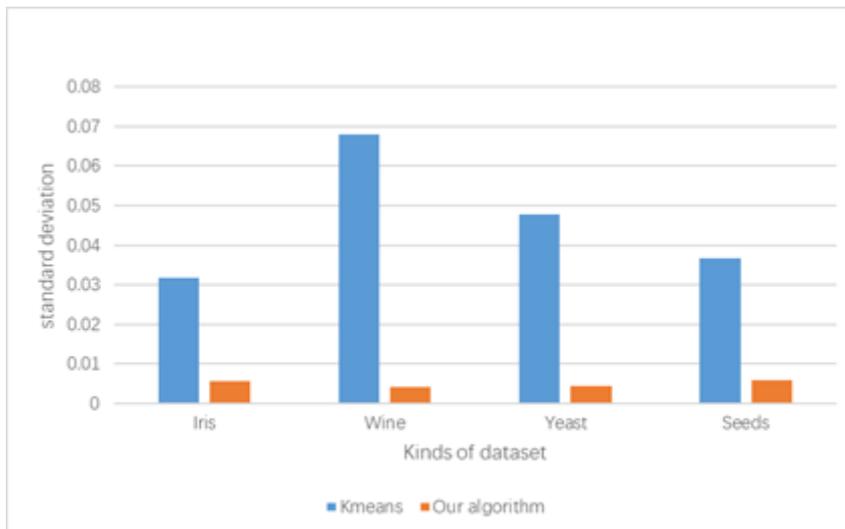

**Figure 3.3 Standard deviations of two algorithms on four datasets**

As the figures clearly depict, our algorithm can create more accurate and robust solutions.

To provide an analysis of the quality of our results relative to the *K*-means algorithm of Spark MLlib, we conduct statistical tests of the hypothesis that the results of the two methods are significantly different. Since the distribution of the results obtained by the algorithms is unknown, a non-parametrical test should be used. Like Ablanedo-Rosas and Rego (2010), we use the Wilcoxon signed rank test to determine whether two samples were selected from populations having the same distribution. The null hypothesis is that the populations of results obtained by *K*-means and our algorithm are identical.

The open source *R* statistical software is used to conduct the test by calling wilcox.test(a,b, paired=TRUE), which is provided with the results from our four datasets listed in Tables 3.2 to

3.5. The *p*-value of this test is: 7.451e-08 and the output from R is: "alternative hypothesis: true location shift is not equal to 0".

Therefore, we reject the null hypothesis and conclude that the results obtained by our algorithm significantly differ from those obtained by the $K$-means algorithm of MLlib. Because the accuracies of our results are better than those obtained by the *K*-means method, statistically our algorithm performs better.

*3.3 Comparison in Accelerating Ratio*

In the field of parallel computing, the *accelerating ratio* is used to indicate how fast the parallel algorithm runs compared to its corresponding sequential execution. The accelerating ratio is defined by:

$$E_r = T_s / T_r \qquad (3.2)$$

In (3.2), $T_s$ represents the time required for the algorithm to run in a traditional sequential manner, and $T_r$ indicates the time required for the algorithm to run in a cluster environment consisting of $r$ computing nodes. A higher accelerating ratio $E_r$ indicates that less computational time is required by the algorithm in a parallel computing environment and thus indicates the higher efficiency of parallelization. In this experiment, we capture the accelerating ratios for our algorithm and the Spark MLlib K-means algorithm using two datasets. The experimental results yield the following figures.

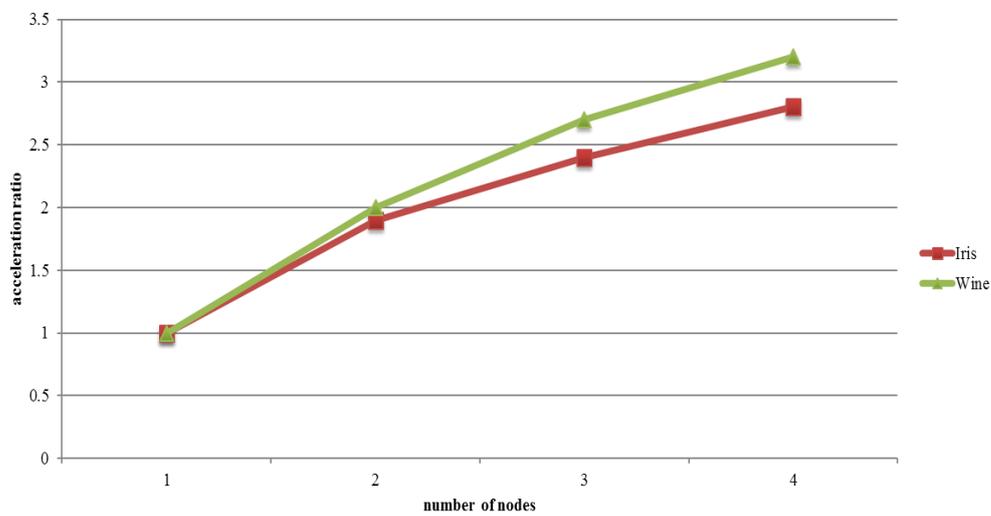

**Figure 3.4 The accelerating ratio of *K*-means in Spark MLlib**

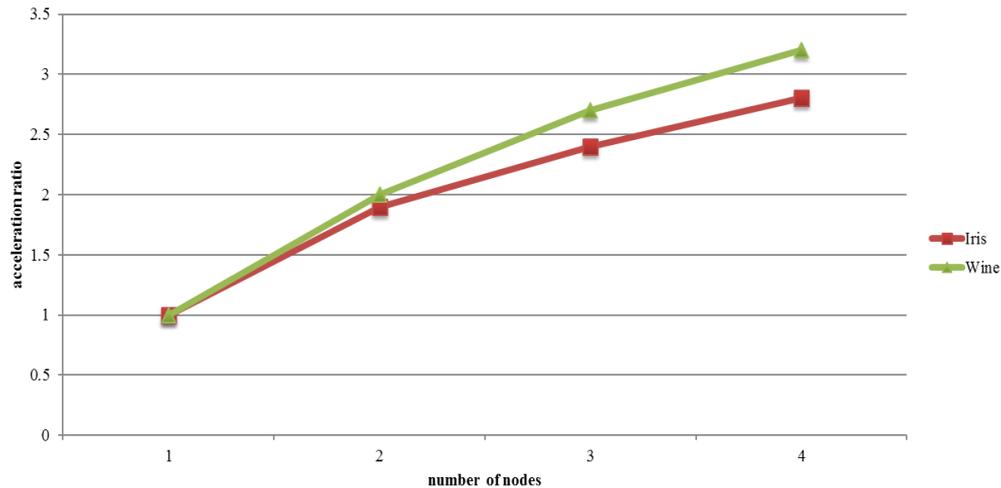

**Figure 3.5 The accelerating ratios of our algorithm in the Spark environment**

Figures 3.4 and 3.5 illustrate the accelerating ratios of the K-means algorithm in Spark MLlib and our algorithm in the Spark environment. Both algorithms are run in parallel mode on multiple computing units/nodes. These figures show that the accelerating ratios of the two algorithms are almost the same, and hence the integration of the Tabu Search component in our algorithm does not negatively impact the accelerating ratio.

On the other hand, the accelerating ratios for both algorithms are not particularly high. This is to be expected given the fact that the sizes of the datasets used in these experiments are not very large (less than 10M). When a data is loaded as an RDD, it can fit completely into the memory of a single computing node. Consequently, the algorithm runs relatively efficiently even in sequential mode when it is run on the single computing node. But when the data is split to be allocated to multiple computing nodes, then as the number of these computing nodes increases, the overhead between nodes also grows due to increased data shuffling. Hence, this counteracts the performance increase brought about by introducing more computing nodes. From the curves in Figures 3.4 and 3.5, we see that the acceleration slows down when the number of computing nodes is greater than 2 for our experiments.

To better validate the speedup of our algorithm, we use some big datasets found in practical logistics applications (Cao and Glover, 2010). Data points are the x-,y-coordinates of customer locations. Formula (3.2) is used to measure the algorithm speed-up. The outcomes are presented in Table 3.6 and Figure 3.6 respectively, where five clusters are built and the parameters for the algorithm are defined as previously stated, i.e., tabu tenure = 5, number of iterations = 1000, and neighborhood size = standard. For the parallel experiments, the configuration consists of one master and three slaves. The solution values are averages over ten runs.

Table 3.6 Computational results for big datasets

| Number of data points | Value of $T_s/T_r$ |
|---|---|
| 931 | 2.8 |
| 1766 | 2.9 |
| 2512 | 2.98 |
| 3180 | 3.15 |
| 3817 | 3.23 |
| 4396 | 3.36 |
| 4936 | 3.35 |
| 5386 | 3.48 |
| 5834 | 3.52 |
| 6257 | 3.55 |
| 6706 | 3.56 |

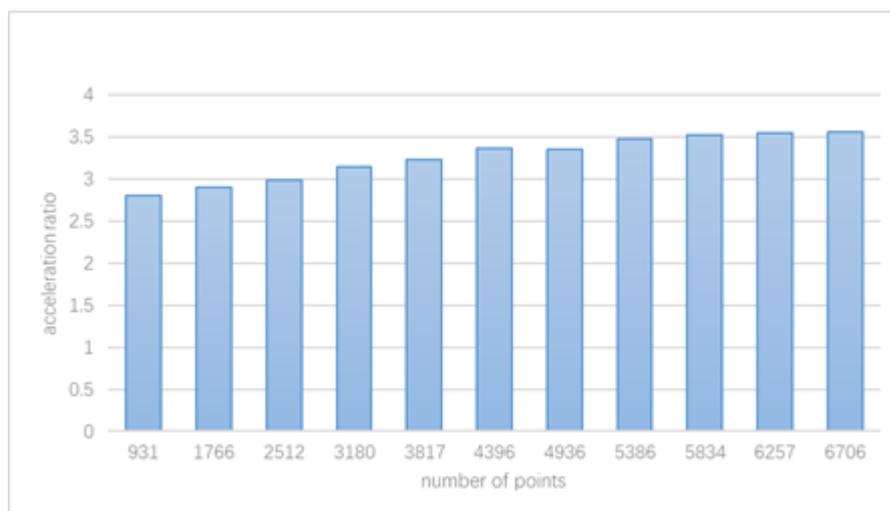

Figure 3.6 The acceleration of our algorithm

As we anticipated that for larger datasets, the effectiveness of parallelization is significant. The parallel version of our algorithm runs three times faster than its sequential counterpart. It should additionally be noted that the *K*-means algorithm in Spark MLlib has been highly tuned and can solve very large-scale clustering problems. According to the computational experiments, the ability of our algorithm to attain a similar accelerating ratio while yielding superior solutions and exhibiting a more robust performance and scalability bodes well for the potential of the parallel version of our algorithm to provide similar advantages for large-scale clustering problems. The fact that Tabu Search typically provides increasingly improved solutions as problem size grows reinforces this expectation.

## 4. CONCLUSIONS

Our Tabu Search based clustering algorithm utilizes the centroid-driven orientation of the $K$-means algorithm under the guidance of a simple version of Tabu Search. Given that the non-centroid data points can be assigned to the proper clusters based on their distances to the centroids without knowing the individual "coordinates" or attributes of each data point, the centroid-driven strategy of our algorithm facilitates its parallel implementation in the Spark environment. One of the primary objectives of our research is to explore the possibility of implementing complicated metaheuristics such as Tabu Search in a Spark environment. To our knowledge, no algorithm based on Tabu Search has previously been implemented on the Spark platform.

Computational experiments disclose that our algorithm can generate better solutions than the *K*-means algorithm of Spark MLLib in terms of both quality and stability, while achieving a similar accelerating ratio when run on multiple computing nodes. These outcomes motivate the exploration of clustering applications from additional settings using Tabu Search by making use of Spark or other big data computing infrastructures.

In future research, we plan to investigate the following enhancements of our approach motivated by the findings reported here:
- add a self-evaluation mechanism to our algorithm which will allow the number of clusters built to be decided automatically.
- revise the implementation of our algorithm to follow the protocol of Spark MLlib, which will ultimately enable our method to be integrated with Spark MLlib for open source.
- incorporate more advanced forms of Tabu Search and more sophisticated neighborhood/candidate list strategies to further improve the efficiency of our algorithm.
- explore opportunities to implement additional metaheuristics in the Spark environment.

**Acknowledgement**: We are indebted to two anonymous reviewers for insightful observations and suggestions that have helped to improve our paper. This work was partially supported by the China Intelligent Urbanization Co-Creation Center [grant number CIUC20150011].